\documentclass[preprint,showpacs,amsmath]{revtex4}
\usepackage{amssymb}
\usepackage{graphicx,epsfig,subfigure,dsfont,amssymb,amsthm,amsfonts,amsbsy,mathrsfs,amscd}
\def\qed{\leavevmode\unskip\penalty9999 \hbox{}\nobreak\hfill
     \quad\hbox{\leavevmode  \hbox to.77778em{%
               \hfil\vrule   \vbox to.675em%
               {\hrule width.6em\vfil\hrule}\vrule\hfil}}
     \par\vskip3pt}

\newtheorem{theorem}{Theorem}
\newtheorem{corollary}{Corollary}

\input amssym.def
\begin{document}

\title{On Coherence of Assistance and Regularized Coherence of Assistance}

\smallskip
\author{Ming-Jing Zhao$^1$}
\author{Teng Ma$^2$ }
\author{Shao-Ming Fei$^{3,4}$ }

\affiliation{
$^1$School of Science,
Beijing Information Science and Technology University, Beijing, 100192, China\\
$^2$State Key Laboratory of Low-Dimensional Quantum Physics and Department of Physics, Tsinghua University, Beijing 100084, China\\
$^3$School of Mathematical Sciences, Capital Normal
University, Beijing
100048, China\\
$^4$Max-Planck-Institute for Mathematics in the Sciences, 04103
Leipzig, Germany}


\begin{abstract}
We study the relation between the coherence of assistance and the regularized coherence of assistance introduced in [Phys. Rev. Lett. {\bf 116}, 070402 (2016)]. The
necessary and sufficient conditions that these two quantities coincide are provided.
Detailed examples are analyzed and the optimal pure state decompositions such that the coherence of assistance equals to the regularized coherence of assistance are derived.
Moreover, we present the protocol for obtaining the maximal relative entropy coherence, assisted by another party under local measurement and one-way communication in one copy setting.
\end{abstract}

\maketitle

\section{Introduction}
Quantum coherence is an important feature in quantum physics \cite{A. Streltsov-rev}. It is also a powerful resource for quantum metrology \cite{V. Giovannetti}, entanglement creation \cite{J. K. Asboth}, and biological processes \cite{E. Collini,N. Lambert,J. Cai,E. J.OReilly}.
Due to the significant roles played in many novel quantum phenomena, it has attracted much attention recently. A rigorous framework for the quantification of coherence is introduced and some intuitive and
computable measures of coherence are identified, for example, the relative entropy coherence and $l_1$ norm coherence  \cite{T. Baumgratz}. The relative entropy coherence of a state is defined as the difference of  von Neumann entropy between the density matrix and the diagonal matrix given by its diagonal entries. The $l_1$ norm coherence depends on the magnitudes of off-diagonal entries of a density matrix.
Trace norm coherence is a coherence measure for qubits \cite{L. H. Shao}, but it is only
a coherence monotone for X states \cite{S. Rana}. Besides, the coherence can also be quantified via the convex roof construction \cite{X. Yuan}.

More than that, there are operational coherence measures such as distillable coherence and
coherence cost which characterize the optimal rate of performance for certain information processing tasks \cite{A. Winter}. In Ref. \cite{A. Winter}, they
reveal the appealing feature of distillable coherence being equal to the relative entropy coherence.
As the maximal average relative entropy coherence of a quantum state, the coherence of assistance $C_a$ is another coherence monotone \cite{E. Chitambar}.
This quantity $C_a$ has an operational interpretation.
Suppose Bob holds a state $\rho^B$. Alice holds another part of the purified state of $\rho^B$. With the help of Alice by performing local measurements and telling Bob her measurement outcomes by classical communication,
the relative entropy coherence of $\rho^B$ can be increased to $C_a(\rho^B)$ maximally.
In many copy setting, if Alice is allowed to make joint measurement across her many copies and telling Bob her measurement results by classical communication,
averagely,
the relative entropy coherence of $\rho^B$ can be increased to $C_a^{\infty}(\rho^B)$, which is called the regularized coherence of assistance \cite{E. Chitambar}.

For the process of increasing relative entropy coherence with the help of another party under the local measurement and one way classical communication, an interesting and meaningful question is when the coherence $C_a$ obtained in one copy setting equals to that $C_a^{\infty}$ in many copy setting.
Obviously, for quantum states $\rho$ such that $C_a(\rho)=C_a^{\infty}(\rho)$, one copy setting is enough, and many copy setting is redundant and wasteful. In this paper, we aim to answer this question and provide analytical results for the equivalence of the coherence of assistance and the regularized coherence of assistance.

First we present the necessary and sufficient conditions when the coherence of assistance attains the regularized coherence of assistance.
Detailed examples are analyzed for two dimensional, three dimensional and high dimensional systems.
In these examples, the optimal decompositions for the saturation of the coherence of assistance $C_a$ with the regularized coherence of assistance $C_a^{\infty}$ are provided. The optimal protocol of obtaining maximal relative entropy coherence assisted by an assistant using local measurement and one way communication in one copy setting is designed finally.

\section{Coherence of assistance}

Under fixed reference basis, the coherence of assistance of a state $\rho$ is characterized by the maximal average relative entropy coherence,
\begin{eqnarray}
C_a(\rho)=\max \sum_i p_i C_r(|\psi_i\rangle),
\end{eqnarray}
where the maximization is taken over all pure state decompositions of $\rho=\sum_i p_i |\psi_i\rangle\langle\psi_i|$, $C_r(\rho)=S(\Delta(\rho))-S(\rho)$ is the relative entropy of coherence,
$\Delta(\rho)$ denotes the state given by the diagonal entries of $\rho$, $S(\rho)$ is the von Neumann entropy \cite{E. Chitambar}.

Coherence of assistance can be interpreted operationally.
For given quantum state $\rho$, its initial relative entropy coherence is $C_r(\rho)$. Now suppose Bob holds a state $\rho^B$ and an assistant Alice holds another part of a purification of $\rho^B$. With the help of Alice by performing local measurement and telling Bob her measurement outcomes by classical communication, the quantum state in Bob will be in one pure state ensemble $\{ p_i,\  |\psi_i\rangle\}$ with relative entropy coherence $\sum_i p_i C_r(|\psi_i\rangle)$. The relative entropy coherence in Bob is increased as relative entropy coherence is monotonic under selective measurements on average. Maximally, the relative entropy coherence can be increased to $C_a(\rho^B)$ in this process.

Similarly,
the regularized coherence of assistance is introduced as the average coherence of assistance in many copy setting,
\begin{eqnarray}\label{def regularized ca}
C_a^{\infty}(\rho)=\lim_{n\to \infty} \frac{1}{n}C_a(\rho^{\otimes n}).
\end{eqnarray}
It is obvious that the coherence of assistance is bounded by the regularized coherence of assistance from above
\begin{eqnarray}\label{relation between ca and cainf}
C_a(\rho)\leq C_a^{\infty}(\rho).
\end{eqnarray}
Utilizing the relation between the regularized coherence of assistance and the regularized entanglement of assistance \cite{D. DiVincenzo, E. Rains}, the authors in \cite{E. Chitambar} have shown a closed form expression for the regularized coherence of assistance,
\begin{eqnarray}\label{exp cainf}
C_a^{\infty}(\rho)=S(\Delta(\rho)).
\end{eqnarray}
Based on this formula, we can get the first necessary and sufficient condition for the saturation of the coherence of assistance with the regularized coherence of assistance as follows.

\begin{theorem}\label{upper bound of ca}
For any quantum state $\rho$, $C_a(\rho)= C_a^{\infty}(\rho)$
if and only if there exists a pure state decomposition
$\rho=\sum_i p_i |\psi_i\rangle\langle\psi_i|$ such that all $\Delta(|\psi_i\rangle)=\Delta(\rho)$.
\end{theorem}

[{Proof}]. By definition, we have
$C_a(\rho)=\max \sum_i p_i C_r(|\psi_i\rangle)
=\max \sum_i p_i S(\Delta(|\psi_i\rangle))
\leq \max  S(\sum_i p_i \Delta(|\psi_i\rangle))
=S(\Delta(\rho))=C_a^{\infty}(\rho)$,
where the second equation is due to $S(|\psi_i\rangle)=0$ for pure state $|\psi_i\rangle$, and the third inequality is from the concavity of the Von Neumann entropy. The third inequality becomes equality if and only if $\Delta(|\psi_i\rangle)$ are the same for all $i$. Hence, the coherence of assistance equals to the regularized coherence of assistance if and only if there exists a pure state decomposition $\rho=\sum_i p_i |\psi_i\rangle\langle\psi_i|$ such that all $\Delta(|\psi_i\rangle)=\Delta(\rho)$, that is all components in the pure state decomposition have the same diagonal entries as the density matrix. \qed

From theorem \ref{upper bound of ca} one can get another necessary and sufficient condition which is easy to prove.

\begin{corollary}\label{th 1'}
For any quantum state $\rho$, $C_a(\rho)= C_a^{\infty}(\rho)$
if and only if there exists a pure state decomposition
$\rho=\sum_i p_i |\psi_i\rangle\langle\psi_i|$ such that each pure state $|\psi_i\rangle$ has relative entropy coherence $S(\Delta(\rho))$.
\end{corollary}

Theorem \ref{upper bound of ca} and corollary \ref{th 1'} are both necessary and sufficient conditions for the coincidence of the coherence of assistance and the regularized coherence of assistance. The former gives more explicit form of the optimal pure state ensemble and the latter is more easy to understand.

$C_a$ is called additive theoretically if $C_a= C_a^{\infty}$. In
Ref. \cite{E. Chitambar} it has been shown that $C_a$ fails to be additive in general, with an example in 4 dimensional system showing the nonadditivity.
Nevertheless, $C_a$ is additive in two dimensional system. Furthermore, we can find one optimal decomposition for the balance of the coherence of assistance and the regularized coherence of assistance by theorem \ref{upper bound of ca}.
Consider two dimensional quantum states
\begin{equation}\label{2-dim state}
\rho=\sum_{i,j=1}^2 \rho_{ij} |i\rangle\langle j|.
\end{equation}
If the coefficient $\rho_{12}$ is real, we choose
\begin{equation}
\begin{array}{rcl}
|\psi_0\rangle&=&\sqrt{\rho_{11}}|1\rangle + \sqrt{\rho_{22}}|2\rangle,\\
|\psi_1\rangle&=&\sqrt{\rho_{11}}|1\rangle - \sqrt{\rho_{22}}|2\rangle,
\end{array}
\end{equation}
and $p_0=\frac{1}{2}(1+\rho_{12}/\sqrt{\rho_{11}\rho_{22}})$, $p_1=\frac{1}{2}(1-\rho_{12}/\sqrt{\rho_{11}\rho_{22}})$ for nonzero $\rho_{11}$ and $\rho_{22}$.
If the coefficient $\rho_{12}$ is complex, with $|\rho_{12}|$ the magnitude and $\arg(\rho_{12})$ the argument, we set
\begin{equation}
\begin{array}{rcl}
|\psi_0\rangle&=&\sqrt{\rho_{11}}|1\rangle + \sqrt{\rho_{22}}e^{-{\rm i}\arg(\rho_{12})}|2\rangle,\\
|\psi_1\rangle&=&\sqrt{\rho_{11}}|1\rangle + \sqrt{\rho_{22}}e^{-{\rm i}(\pi+\arg(\rho_{12}))}|2\rangle,
\end{array}
\end{equation}
and $p_0=\frac{1}{2}(1+|\rho_{12}|/\sqrt{\rho_{11}\rho_{22}})$, $p_1=\frac{1}{2}(1-|\rho_{12}|/\sqrt{\rho_{11}\rho_{22}})$ for nonzero $\rho_{11}$ and $\rho_{22}$.
Thus $\{p_i, |\psi_i\rangle\}$ is an optimal pure state decomposition of $\rho$ such that the coherence of assistance attains the regularized coherence of assistance.
In fact, there are infinitely many optimal decompositions as the choices of the relative phase in $|\psi_0\rangle$ are infinite.
However, once $|\psi_0\rangle$ is fixed, $|\psi_1\rangle$ and the corresponding probabilities $p_0$ and $p_1$ are determined.
Moreover, if one of the elements $\rho_{11}$ and $\rho_{22}$ is zero, the quantum state $\rho$ is pure, and
its coherence of assistance, regularized coherence of assistance and relative entropy coherence are the same.

Now we consider the equality $C_a(\rho)= C_a^{\infty}(\rho)$ in $n$-dimensional systems and investigate the requirement quantum states should satisfy.
For an $n$-dimensional quantum state $\rho=\sum_{ij} \rho_{ij} |i\rangle\langle j|$, we define the matrix equation,
\begin{equation}\label{equation for p}
AP=B.
\end{equation}
Here
\begin{equation}
A=\left(
\begin{array}{ccccccc}
e^{{\rm i}\theta^{(1)}_{{12}}} & e^{{\rm i}\theta^{(2)}_{{12}}} & \cdots & e^{{\rm i}\theta^{(T)}_{{12}}}\\
e^{{\rm i}\theta^{(1)}_{{13}}} & e^{{\rm i}\theta^{(2)}_{{13}}} & \cdots & e^{{\rm i}\theta^{(T)}_{{13}}}\\
\cdots & \cdots & \cdots & \cdots \\
e^{{\rm i}\theta^{(1)}_{{1n}}} & e^{{\rm i}\theta^{(2)}_{{1n}}} & \cdots & e^{{\rm i}\theta^{(T)}_{{1n}}}\\
e^{{\rm i}\theta^{(1)}_{{23}}} & e^{{\rm i}\theta^{(2)}_{{23}}} & \cdots & e^{{\rm i}\theta^{(T)}_{{23}}}\\
\cdots & \cdots & \cdots & \cdots \\
e^{{\rm i}\theta^{(1)}_{{2n}}} & e^{{\rm i}\theta^{(2)}_{{2n}}} & \cdots & e^{{\rm i}\theta^{(T)}_{{2n}}}\\
\cdots & \cdots & \cdots & \cdots \\
e^{{\rm i}\theta^{(1)}_{{n-1,n}}} & e^{{\rm i}\theta^{(2)}_{{n-1,n}}} & \cdots & e^{{\rm i}\theta^{(T)}_{{n-1,n}}}\\
1 & 1 & \cdots & 1
\end{array}
\right)_{(\frac{n(n-1)}{2}+1)\times T}
\end{equation}
with ${(n-1)(n-2)}/{2}$ constraints
\begin{equation}\label{theta constrains}
\left\{\begin{array}{rcl}
\theta^{(k)}_{1s}-\theta^{(k)}_{2s}&=&\theta^{(k)}_{12},\ \ s=3,\cdots,n,\\
\theta^{(k)}_{2s}-\theta^{(k)}_{3s}&=&\theta^{(k)}_{23},\ \ s=4,\cdots,n,\\
\cdots\\
\theta^{(k)}_{n-2,n}-\theta^{(k)}_{n-1,n}&=&\theta^{(k)}_{n-2,n-1},
\end{array}
\right.
\end{equation}
for all $k$. There are essentially $n-1$ independent variables $\theta^{(k)}_{ij}$ for each $k$, $k=1,2,\cdots,T$, which are all between 0 and $2\pi$. $P=(p_1,p_2,\cdots,p_T)^t$, $0\leq p_k\leq 1$ for $k=1,2...,T$.
\begin{small}
\begin{equation}
B=(\frac{\rho_{12}}{\sqrt{\rho_{11}\rho_{22}}}, \frac{\rho_{13}}{\sqrt{\rho_{11}\rho_{33}}},\cdots,\frac{\rho_{1n}}{\sqrt{\rho_{11}\rho_{nn}}},\frac{\rho_{23}}{\sqrt{\rho_{22}\rho_{33}}}, \cdots,
\frac{\rho_{2n}}{\sqrt{\rho_{22}\rho_{nn}}},\cdots,\frac{\rho_{n-1,n}}{\sqrt{\rho_{n-1,n-1}\rho_{nn}}},1)^t,
\end{equation}
\end{small}
with superscript $t$ denoting transpose.
For vector $B$, although its components are all fractions,  if one denominator is zero, then the corresponding numerator must be zero because of the positivity of density matrix. Therefore, vector $B$ is a well defined $\frac{n(n-1)}{2}+1$ dimensional vector and is decided by the coefficients of density matrix. In Eq. (\ref{equation for p}), the vector $B$ is known and given by the density matrix, the
matrix $A$ and the vector $P$ are unknown.

\begin{theorem}\label{th n if and only if}
For $n$-dimensional quantum state $\rho$, $C_a(\rho)=C_a^{\infty}(\rho)$ if and only if the equation (\ref{equation for p}) has solutions for unknowns $P$ and $\theta^{(k)}_{{ij}}$ satisfying conditions (\ref{theta constrains}).
\end{theorem}

[{Proof}].
Let $\{p_k, |\psi_k\rangle\}$
be an optimal pure state ensemble such that $C_a(\rho)=\sum_{k=1}^T p_k C_r(|\psi_k\rangle\langle\psi_k|)$. If $C_a(\rho)$ attains its upper bound $C_a^{\infty}(\rho)$, then $C_r(|\psi_k\rangle\langle\psi_k|)=S(\Delta(\rho))$ by corollary \ref{th 1'} and
$|\psi_k\rangle\langle\psi_k|$ should be of the form
\begin{equation}\label{eq n pure}
\left(
\begin{array}{cccccc}
\rho_{11}& \sqrt{\rho_{11}\rho_{22}}e^{{\rm i}\theta^{(k)}_{{12}}} & \sqrt{\rho_{11}\rho_{33}}e^{{\rm i}\theta^{(k)}_{{13}}}  & \cdots
& \sqrt{\rho_{11}\rho_{nn}}e^{{\rm i}\theta^{(k)}_{{1n}}}\\
\sqrt{\rho_{11}\rho_{22}}e^{{\rm -i}\theta^{(k)}_{{12}}} & \rho_{22} & \sqrt{\rho_{22}\rho_{33}}e^{{\rm i}\theta^{(k)}_{{23}}} & \cdots
& \sqrt{\rho_{22}\rho_{nn}}e^{{\rm i}\theta^{(k)}_{{2n}}}\\
\sqrt{\rho_{11}\rho_{33}}e^{{\rm -i}\theta^{(k)}_{{13}}}  & \sqrt{\rho_{22}\rho_{33}}e^{{-\rm i}\theta^{(k)}_{{23}}} & \rho_{33}& \cdots
& \sqrt{\rho_{33}\rho_{n,n}}e^{{\rm i}\theta^{(k)}_{{3n}}}\\
\cdots &\cdots &\cdots &\cdots &\cdots\\
\sqrt{\rho_{11}\rho_{nn}}e^{{\rm -i}\theta^{(k)}_{{1n}}} &  \sqrt{\rho_{22}\rho_{nn}}e^{{\rm -i}\theta^{(k)}_{{2n}}}
&\sqrt{\rho_{33}\rho_{nn}}e^{{\rm -i}\theta^{(k)}_{{3n}}} & \cdots
&\rho_{nn}
\end{array}
\right)
\end{equation}
by theorem \ref{upper bound of ca} for all $k$.
The ${(n-1)(n-2)}/{2}$ constraints in Eqs. (\ref{theta constrains})
guarantee that the rank of $|\psi_k\rangle\langle\psi_k|$ in Eq. (\ref{eq n pure}) is one.
$\rho=\sum_{k=1}^T p_k |\psi_k\rangle\langle\psi_k|$ demands
\begin{equation}
\sum_{k=1}^T p_k \sqrt{\rho_{ii}\rho_{jj}}e^{{\rm i}\theta^{(k)}_{{ij}}}=\rho_{ij},
\end{equation}
or
\begin{equation}
\sum_{k=1}^T p_k e^{{\rm i}\theta^{(k)}_{{ij}}}=\rho_{ij}/\sqrt{\rho_{ii}\rho_{jj}},
\end{equation}
for $1\leq i<j\leq n$, which gives rise to the equation (\ref{equation for p}). Thus $C_a(\rho)=C_a^{\infty}(\rho)$ if and only if the equation (\ref{equation for p}) has solutions for $P$ and $\theta^{(k)}_{{ij}}$ satisfying conditions (\ref{theta constrains}).
\qed

In Theorem 1 and Corollary 1, the necessary and sufficient conditions are provided for the saturation of the coherence of assistance $C_a(\rho)$ with the regularized coherence of assistance $C_a^{\infty}(\rho)$.
In theorem 2, we present the way to find the optimal pure state ensemble for this saturation. The solution $P$ in matrix equation (8)
is just the probabilities $\{p_k\}$ in the optimal decomposition $\{p_k, |\psi_k\rangle\}$. The solution $\theta^{(k)}_{{ij}}$ in $A$ in (8) is the argument of the entries in the $i$-th row and the
$j$-th column with magnitude $\sqrt{\rho_{ii}\rho_{jj}}$ for the component $|\psi_k\rangle\langle\psi_k|$ in the optimal decomposition $\{p_k, |\psi_k\rangle\}$. The problem of theorem 2
is that the matrix $A$ and $P$ scale quadratically with respect to the dimension of the density matrix, which implies
more unknowns $P$ and arguments $\theta$ in $A$ are involved when the dimension increases. In solving the matrix equation,
one can select proper independent arguments first, then subsequently the matrix $A$.
The vector $P$ is then determined by $A$ and the previous vector $B$.
If $P=(p_1,p_2,\cdots,p_T)^t$ is the solution satisfying $0\leq p_k\leq 1$ for $k=1,2...,T$, then the solution is obtained
and the coherence of assistance $C_a(\rho)$ equals to regularized coherence of assistance $C_a^{\infty}(\rho)$.
Otherwise, one chooses different independent arguments.

{\it Example 1}. Consider the following three dimensional state,
\begin{equation}\label{3-dim state}
\rho=\sum_{i,j=1}^3 \rho_{ij} |i\rangle\langle j|.
\end{equation}
According to Theorem 2, $C_a(\rho)=C_a^{\infty}(\rho)$ if and only if matrix equation
\begin{equation}\label{eq dim 3}
\left(
\begin{array}{ccccccc}
e^{{\rm i}\theta^{(0)}_{{12}}} & e^{{\rm i}\theta^{(1)}_{{12}}} & \cdots & e^{{\rm i}\theta^{(T-1)}_{{12}}}\\
e^{{\rm i}\theta^{(0)}_{{23}}} & e^{{\rm i}\theta^{(1)}_{{23}}} & \cdots & e^{{\rm i}\theta^{(T-1)}_{{23}}}\\
e^{{\rm i}\theta^{(0)}_{{13}}} & e^{{\rm i}\theta^{(1)}_{{13}}} & \cdots & e^{{\rm i}\theta^{(T-1)}_{{13}}}\\
1 & 1 & \cdots & 1
\end{array}
\right)
\left(
\begin{array}{ccccccc}
p_0\\p_1\\\cdots\\p_{T-1}
\end{array}
\right)
=\left(
\begin{array}{ccccccc}
\rho_{12}/\sqrt{\rho_{11}\rho_{22}}\\
\rho_{23}/\sqrt{\rho_{22}\rho_{33}}\\
\rho_{13}/\sqrt{\rho_{11}\rho_{33}}\\
1
\end{array}
\right)
\end{equation}
with $\theta^{(k)}_{{12}}+\theta^{(k)}_{{23}}=\theta^{(k)}_{{13}}$,
have solutions for $P$ satisfying $0\leq p_k\leq 1$ and free arguments $\theta^{(k)}_{{12}}$ and $\theta^{(k)}_{{23}}$.

For simplicity, suppose $\rho_{12}$, $\rho_{23}$ and $\rho_{13}$ are all non-zero real numbers.
Denote $\rho_{12}/\sqrt{\rho_{11}\rho_{22}}=r_1$, $\rho_{23}/\sqrt{\rho_{22}\rho_{33}}=r_2$ and $\rho_{13}/\sqrt{\rho_{11}\rho_{33}}=r_3$.
First, set $T=4$ and $\theta^{(0)}_{{12}}=\theta^{(0)}_{{23}}=0$, $\theta^{(1)}_{{12}}=\pi$, $\theta^{(1)}_{{23}}=0$, $\theta^{(2)}_{{12}}=\theta^{(2)}_{{23}}=\pi$, $\theta^{(3)}_{{12}}=0$, $\theta^{(3)}_{{23}}=\pi$.
Then the matrix equation (\ref{eq dim 3}) becomes
\begin{equation}
\left(
\begin{array}{ccccccc}
1 & -1 & -1 & 1\\
1 & 1 & -1 & -1\\
1& -1 & 1 & -1\\
1 & 1 & 1 & 1
\end{array}
\right)
\left(
\begin{array}{ccccccc}
p_0\\p_1\\p_2\\p_3
\end{array}
\right)
=\left(
\begin{array}{ccccccc}
r_1\\
r_2\\
r_3\\
1
\end{array}
\right).
\end{equation}
The unique solution of the matrix equation above is $p_0=\frac{1}{4}(r_1+r_2+r_3+1)$, $p_1=\frac{1}{4}(r_2-r_1-r_3+1)$, $p_2=\frac{1}{4}(r_3-r_1-r_2+1)$, $p_3=\frac{1}{4}(r_1-r_2-r_3+1)$.
Obviously, $p_0,p_1,p_2,p_3\leq 1$. Therefore, if $r_1+r_2+r_3+1\geq 0$, $r_1-r_2-r_3+1\geq 0$, $r_2-r_1-r_3+1\geq 0$ and $r_3-r_1-r_2+1\geq 0$, then $\{p_i\}$ and $\{\theta_{ij}^{(k)}\}$ are one set of solutions of Eq. (\ref{eq dim 3}) for $C_a(\rho)=C_a^{\infty}(\rho)$. Therefore the probabilities $\{p_i\}$ with pure states
\begin{equation}\label{3-dim optimal dec}
\begin{array}{rcl}
|\psi_0\rangle&=&\sqrt{\rho_{11}}|1\rangle+\sqrt{\rho_{22}}|2\rangle+\sqrt{\rho_{33}}|3\rangle, \\ |\psi_1\rangle&=&-\sqrt{\rho_{11}}|1\rangle+\sqrt{\rho_{22}}|2\rangle+\sqrt{\rho_{33}}|3\rangle,\\
|\psi_2\rangle&=&\sqrt{\rho_{11}}|1\rangle-\sqrt{\rho_{22}}|2\rangle+\sqrt{\rho_{33}}|3\rangle,\\
|\psi_3\rangle&=&\sqrt{\rho_{11}}|1\rangle+\sqrt{\rho_{22}}|2\rangle-\sqrt{\rho_{33}}|3\rangle.
\end{array}
\end{equation}
constitute the optimal decomposition of $\rho$ in Eq. (\ref{3-dim state}) giving $C_a(\rho)=C_a^{\infty}(\rho)$.
Such quantum states all belongs to the polyhedron in Fig. 1.

\begin{center}
\begin{figure}[!h]\label{fig}
\resizebox{8cm}{!}{\includegraphics{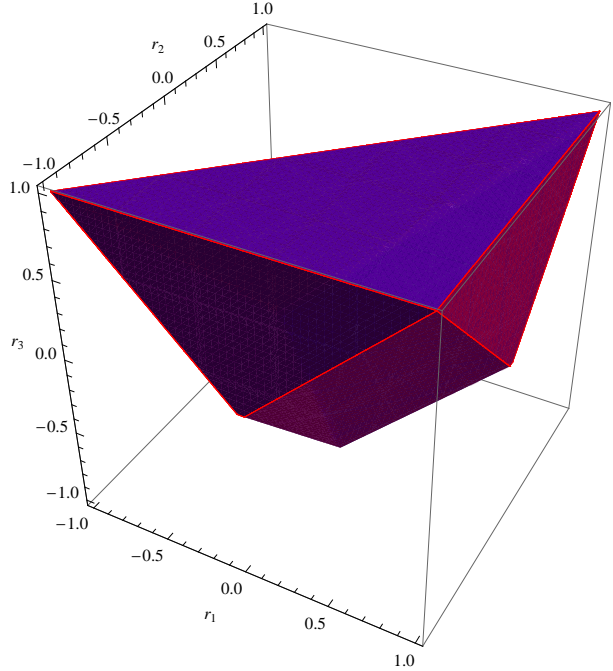}}
\caption{(Color online) Quantum states in this polyhedron satisfy four inequalities: $r_1+r_2+r_3+1\geq 0$, $r_1-r_2-r_3+1\geq 0$, $r_2-r_1-r_3+1\geq 0$ and $r_3-r_1-r_2+1\geq 0$. The coherence of assistance
attains the regularized coherence of assistance for these quantum states.}
\end{figure}
\end{center}

{\it Example 2}. Consider an $n$-dimensional state $\rho=\sum_{i,j=1}^n \rho_{ij} |i\rangle \langle j|$
such that
\begin{eqnarray}\label{eq n-dim}
\sum_{k=1}^{n-1} p_k f(k)+p_0=\rho_{ij}/\sqrt{\rho_{ii}\rho_{jj}}, \ \ i<j,
\end{eqnarray}
holds for some probabilities $p_k$, where $f(k)=1$ for $i\leq k<j$, and $f(k)=-1$ otherwise, $0\leq p_k\leq 1$ for $k=0,1,\cdots,n-1$.
Eq. (\ref{eq n-dim}) is derived by inserting Eq. (\ref{equation for p}) with
$\theta_{1j}^{(1)}=\pi$, $j=2,\cdots, n$; $\theta_{1j}^{(2)}=\theta_{2j}^{(2)}=\pi$, $j=3,\cdots, n$;
$\cdots$;
$\theta_{1n}^{(n-1)}=\cdots=\theta_{nn}^{(n-1)}=\pi$; and other arguments 0.
Therefore if $n$ dimensional quantum state $\rho$ satisfies Eq. (\ref{eq n-dim}) for some probabilities $p_k$, then it allows solution for Eq. (\ref{equation for p}) for some probabilities $p_k$ and $\theta$ defined above. Such quantum state $\rho$ satisfying Eq. (\ref{eq n-dim}) makes $C_a(\rho)=C_a^{\infty}(\rho)$.
For the given arguments
$\theta_{ij}^{(k)}$, we find the corresponding pure states are
\begin{equation}\label{n-dim optimal dec}
\begin{array}{rcl}
|\psi_0\rangle&=&\sqrt{\rho_{11}}|1\rangle+\sqrt{\rho_{22}}|2\rangle+\cdots+\sqrt{\rho_{nn}}|n\rangle,\\ |\psi_1\rangle&=&-\sqrt{\rho_{11}}|1\rangle+\sqrt{\rho_{22}}|2\rangle+\cdots+\sqrt{\rho_{nn}}|n\rangle, \\ |\psi_2\rangle&=&-\sqrt{\rho_{11}}|1\rangle-\sqrt{\rho_{22}}|2\rangle+\cdots+\sqrt{\rho_{nn}}|n\rangle,\\ &\cdots&\\
|\psi_{n-1}\rangle&=&-\sqrt{\rho_{11}}|1\rangle-\sqrt{\rho_{22}}|2\rangle+\cdots-\sqrt{\rho_{n-1,n-1}}|n-1\rangle+\sqrt{\rho_{nn}}|n\rangle.
\end{array}
\end{equation}
Then $\{p_k, |\psi_k\rangle\}$ constitutes an optimal decomposition of $\rho$ with $P=(p_0,p_1,\cdots,p_n)^t$ the solution of Eq. (\ref{eq n-dim}) and $\{|\psi_k\rangle\}$ in Eqs. (\ref{n-dim optimal dec}).

As coherence of assistance $C_a(\rho)$ is the maximal relative entropy coherence obtained with the help of another party making local measurement and one way classical communication in one copy setting. It can be increased more generally in many copy setting. For quantum state $\rho$, the equality $C_a(\rho)=C_a^{\infty}(\rho)$ means to increase the relative entropy coherence in one copy setting is equivalent to the result in many copy setting. Therefore, many copy setting and joint measurement of assistant is redundant.

By theorem 2 we have presented some classes of quantum states whose coherence of assistance $C_a(\rho)$ reaches regularized coherence of assistance $C_a^{\infty}(\rho)$, together with the corresponding optimal pure state decompositions for each class of quantum states. Based on these results, the protocol of obtaining the maximal relative entropy coherence with the help of assistant using local measurement and one way communication can be schemed explicitly.
As an example let us consider the three dimensional quantum state given by Eq. (\ref{3-dim state}), denoted as $\rho_B$, which is held by Bob. As a purification
we first prepare a pure entangled state $|\psi\rangle_{AB}=\sum_{i=0}^3 |i\rangle_A|\psi_i\rangle_B$, with $\{|\psi_i\rangle\}_{i=0}^3$ given in Eqs. (\ref{3-dim optimal dec}). Then Alice performs optimal von Neumann measurements on the basis $\{|i\rangle_A\}$.
If Alice's part is projected to state $|i\rangle_A$, the state of Bob will be collapsed to $|\psi_i\rangle_B$, with relative entropy coherence $S(\Delta(\rho_B))$.
After receiving Alice's measurement outcomes via
classical communication channel, Bob can obtain his state in a four-state ensemble that each state has the same relative entropy coherence $S(\Delta(\rho_B))$. Therefore the final relative entropy coherence for Bob is $S(\Delta(\rho_B))$, which is the maximal relative entropy coherence he can get in this one way assisted protocol.

\section{Conclusions}

To summarize, we have investigated the saturation of the coherence of assistance $C_a(\rho)$ with its upper bound regularized coherence of assistance $C_a^{\infty}(\rho)$. Necessary and sufficient conditions have been provided. Especially, for some special quantum states in two dimensional, three dimensional and general high dimensional systems, the optimal decompositions for the coincidence of $C_a(\rho)$ and $C_a^{\infty}(\rho)$ have been
presented. And the corresponding optimal protocol of obtaining the maximal relative entropy coherence with the help of assistant using local measurement and one way communication has been schemed.
These results are of significant implications in two folds. Firstly, the equality $C_a(\rho)=C_a^{\infty}(\rho)$ implies the additivity of coherence of assistance $C_a(\rho)$. We have investigated which kind of quantum states allow the coherence of assistance additive mathematically. Secondly,
the equality $C_a(\rho)=C_a^{\infty}(\rho)$ shows the equivalence of the maximal relative entropy coherence in one way assisted protocol in one copy setting and that in many copy setting. Here we have revealed the conditions for which kind of quantum states
the maximal relative entropy coherence obtained in one way assisted protocol
with one copy setting is enough.

Note that coherence of assistance $C_a(\rho)$ is the maximal relative entropy coherence attained with the help of another part by local measurements and one way
communication in one copy setting, while the relative entropy coherence is in fact the distillable coherence. Therefore, coherence of assistance $C_a$ quantifies
the one way coherence distillation rate with the help of another part in one copy setting.
In many copy setting, higher one way coherence distillation rate can be obtained. In average
$C_a^{\infty}(\rho)$ characterizes the one way coherence distillation rate in infinite copy setting.
The equality $C_a(\rho)=C_a^{\infty}(\rho)$ shows the equivalence of one way distillation rate in one copy setting and the one way distillation rate in many copy setting assisted by another party.
In Ref. \cite{K. D. Wu}, an experimental realization in linear optical system for obtaining the maximal relative entropy coherence for two dimensional quantum states in assisted distillation
protocol has been presented. Their results are based on one copy setting as the optimal distillable rate of two dimensional quantum states can be reached with one copy scenario. Our research may help for assisted distillation of coherence in high dimensional systems experimentally.

\bigskip
\noindent{\bf Acknowledgments}\, This work is supported by the NSF of China under
Grant Nos. 11401032 and 11675113.

\end{document}